\documentclass[sigplan,9pt,screen,natbib,nonacm]{acmart}
\setcopyright{none}
\usepackage{mdwtab}
\usepackage{mdwlist}
\usepackage{amsmath}
\usepackage{comment}
\usepackage{url}
\usepackage{hyphenat}

\usepackage{listings}

\lstdefinestyle{sOcaml}{language=[Objective]Caml,
  morekeywords={effect,perform,locus},
  literate={+}{{$+$}}1 {/}{{$/$}}1 
           {=}{{$=$}}1
           {>}{{$>$}}1 {<}{{$<$}}1
           {<>}{$\not=$}1
           {->}{{$\rightarrow$}}2 {>=}{{$\geq$}}2 {<-}{{$\leftarrow$}}2
           {<=}{{$\leq$}}2
           {==>}{{$\mapsto$}}2
           {|}{{$\mid$}}1
           {|>}{{$\triangleright$}}1
           {>>}{{$\rhd$}}1
           {'a}{$\alpha$}1
           {'b}{$\beta$}1
           {'c}{$\gamma$}1
           {'e}{$\epsilon$}1
           {'n}{$\nu$}1
           {'w}{$\omega$}1
           {'r}{$\rho$}1
           {'state}{$\sigma$}1
           {'w.}{$\forall\omega.\ $}2
           {:=}{\ensuremath{\mathrel{{:}{=}}}}2
           {...}{\ldots}2
           {\#\#+}{\color{red}}1
           {\#\#-}{\color{black}}1
           {\#\#\#}{{$\leadsto$}}3
}
\lstnewenvironment{code}{\lstset{basicstyle={\linespread{1.02}\sffamily\normalsize}}}{}

\lstMakeShortInline[columns=fullflexible]|

\newcommand{\aside}[1]{\ignorespaces}

\begin{document}

\title{Highest-performance Stream Processing}
\author{Oleg Kiselyov}
\orcid{0000-0002-2570-2186}
\affiliation{%
  \institution{Tohoku University}
  \country{Japan}}
\email{oleg@okmij.org}

\author{Tomoaki Kobayashi}
\affiliation{%
  \institution{Tohoku University}
  \country{Japan}}
\email{tomoaki.kobayashi.t3@dc.tohoku.ac.jp}

\author{Aggelos Biboudis}
\affiliation{%
  \institution{}
  \country{Switzerland}}
\email{biboudis@gmail.com}

\author{Nick Palladinos}
\affiliation{%
  \institution{Nessos IT}
  \country{Greece}}
\email{npal@nessos.gr}

\begin{abstract}
We present the stream processing library that achieves the highest
performance of existing OCaml streaming libraries,
attaining the speed and memory efficiency of hand-written state
machines.  It supports finite and infinite streams with the familiar
declarative interface, of \emph{any} combination of
map, filter, take(while), drop(while), zip, flatmap combinators and
tupling. Experienced users may use the lower-level interface of
stateful streams and implement accumulating maps, compression and windowing.
The library is based on assured code generation (at
present, of OCaml and C) and guarantees in all cases \emph{complete
  fusion}.
\end{abstract}
\maketitle

\section{Summary}
\label{s:summary}
Strymonas is a DSL that generates high-performance single-core stream
processing code from declarative descriptions of stream pipelines and
user actions~-- something like Yacc. Unlike (ocaml)yacc, strymonas is
an embedded DSL. Therefore, it integrates as is with the existing OCaml code and
tools. Any typing or pipeline mis-assembling errors are reported
immediately (even during editing).

Strymonas statically guarantees \emph{complete fusion}: if each
operation in a pipeline individually runs without any function calls
and memory allocations, the entire streaming
pipeline runs without calls and allocations. Thus strymonas per se
introduces not even constant-size intermediary data structures.
Complete fusion is mainly the space guarantee: the ability to run the
processing loop without any GC or even stack allocations. Still,
avoiding closures and the repeated construction/disposal of tuples,
option values, etc. notably improves performance, in our experience.

The present strymonas is the completely re-written and the much
extended and improved version of the library described
\cite{strymonas-2017}. The main differences are:
\begin{itemize}
\item Generated code is not only OCaml but also C. The latter
needs no OCaml runtime and can be automatically vectorized. 
\item The core
of strymonas~-- stream representation and fusion~-- is now pure OCaml,
with code generation relegated to a backend. MetaOCaml is needed only for the
OCaml-code--generation backend~-- but not for the C backend. Users may develop
their own backends by implementing a specific signature.
\item Stream fusion is now achieved in all cases. In
  \cite{strymonas-2017}, complicated zipping pipelines
  did not fuse completely.
\item To guarantee the complete fusion in all cases, including
  complicated zip pipelines, the library was completely re-written,
  using a different fusion approach based on
  normalization-by-evaluation.
\item The library is structured as the low-level core interface of
  stafeful streams upon which the familiar declarative combinators are
  implemented. The core interface can be used directly, to write
  stream processing that depends not only on the current but past
  stream elements~-- such as signal-processing filtering, windowing,
  compression. One may freely mix the two interfaces in the same
  pipeline.
\item The low-level interface supports streams not only over base
  types (integers, strings, floats) but also over tuples, records and even
  abstract data types. The example of the latter is windowing.
\end{itemize}

\subsection{Backends}

Semantic actions~-- stream mapping, filtering, accumulating,
etc. functions~-- are expressed in another embedded DSL, called a
backend. Strymonas currently provides two implementations of the DSL
(with the identical interface): OCaml and C.\footnote{Actually, there
  are four backends: basic OCaml, basic C, and the backends obtained
  by applying a partial-evaluation functor to any backend.} We are
considering WASM and LLVM IR backends.

The OCaml backend is based on MetaOCaml and extends the common backend
interface to permit arbitrary OCaml code (enclosed in MetaOCaml
brackets) as semantic actions.\footnote{If such an extensibility is not
  needed, the OCaml backend can be implemented in pure OCaml: that is,
  without MetaOCaml.} On the other hand, the C backend is pure
OCaml: MetaOCaml installation is \emph{not} needed. No other
dependencies are needed either. The C backend uses the
tagless-final--based approach described in \cite{generating-C}.
Thanks to tagless-final, the backends are extensible.  Either
backend generates code that is statically assured to compile without
errors or warnings.

\section{A Taste of Strymonas}

Strymonas may be thought of as Yacc for stream processing~-- but
embedded rather than standalone.  Here is the simplest example (for
clarity, we explicitly write type annotations although none are
needed):
\begin{code}
let ex1 : int cstream = iota C.(int 1) |> map C.(fun e -> e * e)
\end{code}
Like Yacc, strymonas uses two languages: one to describe the structure
of the stream pipeline, and the other to specify the semantic actions
such as mapping transformations, etc. Since strymonas is an embedded
DSL, both languages are represented by OCaml functions (combinators),
but from two different namespaces (signatures). Stream structure
combinators such as |iota| and |map| produce, consume, or transform
values of the type |'a cstream|, where |'a| is a base
type.\footnote{In a lower-level strymonas interface, stream elements
  are not restricted to base types.} The |iota| combinator produces an
infinite stream of numbers starting with the given one; |map| should
be self-explanatory.  The combinators are typically composed via
\lstinline{|>}: the right-to-left application operator.  Semantic
actions (i.e., the arguments of stream combinators) are described via
backend combinators, which build values of the abstract type 
|'a cde| representing the target code: C, OCaml, etc.\footnote{Thus
  \textsf{$\alpha$ cde} is an abstract, backend-independent
  representation of the generated code, which may not even be an OCaml
  code~-- and hence different from MetaOCaml's \textsf{$\alpha$
    code}. MetaOCaml may not be needed, depending on the
  chosen backend.} We shall assume one such backend in scope, as the
module named |C|.

The |ex1| pipeline is the infinite stream of natural numbers
transformed by squaring each item.  The pipeline is completed by
terminating it by an |int cstream| consumer, such as |fold|, or its
instance |sum| below (strymonas already provides this
   combinator, under the name |sum_int|):
\begin{code}
let sum : int cstream -> int cde = fold C.(+) C.(int 0)
let ex2 : int cde = 
  ex1 |> filter C.(fun e -> e mod (int 17) > int 7)
       |> take C.(int 10) |> sum
\end{code}
We added to |ex1| two more transformations, to retain only those
items whose remainder mod 17 exceeds 7, and take first 10 such
items. The resulting stream becomes finite and can be meaningfully
summed up. Here |ex2| is the |int cde| value representing the
integer-producing target code for the complete pipeline, including
stream generation, folding, and the user actions of squaring, etc. The
backend that realizes |'a cde| as OCaml code lets us pretty-print this
code:
\begin{code}
let v_1 = ref 0 in let v_2 = ref 10 in let v_3 = ref 1 in
while (! v_2) > 0 do
   let t_4 = ! v_3 in incr v_3;
   let t_5 = t_4 * t_4 in
   if (t_5 mod 17) > 7 then (decr v_2; v_1 := ! v_1 + t_5)
done; 
! v_1
\end{code}
The code can also be saved into a file, compiled,
put into a library~-- or
it can be run right away:
dynamically linked into the program that generated it and invoked.
With the C back-end, the
resulting code is (some newlines are removed for compactness):
\begin{code}
int fn()
{ int v_1 = 0; int v_2 = 10; int v_3 = 1;
  while (v_2 > 0)
  { int t_4; int t_5;
    t_4 = v_3;
    v_3++; t_5 = t_4 * t_4;
    if ((t_5 % 17) > 7) { v_2--; v_1 = v_1 + t_5; }
  }
  return v_1;}
\end{code}
This is what a competent programmer would have written by hand.
Although the pipeline is purely declarative, with first-class (the argument of
|filter|) and higher-order functions, the generated code is
imperative and has no function calls. The main loops runs with no GC,
even in OCaml.

The second example is the pipeline to compute the dot-product of two arrays:
\begin{code}
let ex_dot (arr1:int array cde,arr2:int array cde) : int cde = 
  zip_with C.( * ) (of_arr arr1) (of_arr arr2) |> sum
\end{code}
The combinator |of_arr| creates a finite stream whose contents is the
given target language array. With the OCaml backend we generate:
\begin{code}
fun (arg1_24, arg2_25) ->
  let t_26 = (Array.length arg2_25) - 1 in
  let t_27 = (Array.length arg1_24) - 1 in
  let v_28 = ref 0 in
  for i_29 = 0 to if t_27 < t_26 then t_27 else t_26 do
     let el_30 = Array.get arg1_24 i_29 in
     let el_31 = Array.get arg2_25 i_29 in
     v_28 := (! v_28) + (el_30 * el_31)
  done;
  ! v_28
\end{code}
Incidentally, |zip_with f| appearing in |ex_dot| is defined in
strymonas as |zip >> map (fun (x,y) -> f x y)| where |>>| is
left-to-right function composition. A naive implementation would
construct a tuple in |zip|, to be deconstructed in the subsequent
mapping. The strymonas-generated code however clearly has no
tuples.

\section{Evaluation}

\begin{figure*}
\includegraphics[width=0.97\linewidth]{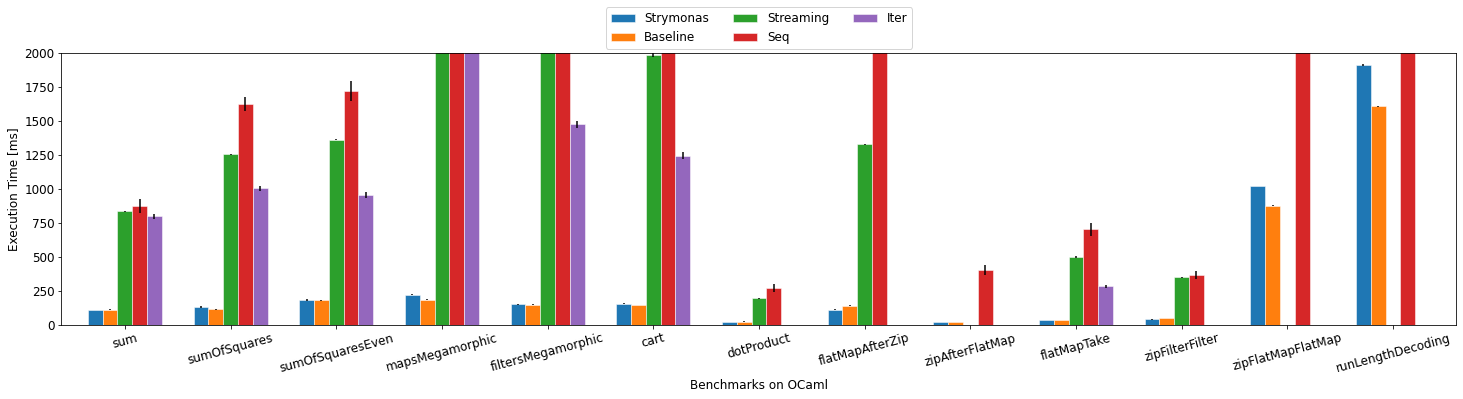}
\caption{Benchmarking against the baseline, `Seq', `iter', and
`streaming': the running time
in milliseconds per iteration (avg. of 20, with
  mean-error bars shown). Shorter bars are better. To better show the details,
  the figure is truncated: the per-iteration running time of 
  Seq on mapsMegamorphic is 7 sec, filtersMegamorphic 4 sec,
  zipFlatMapFlatMap 14 sec and runLengthDecoding 37 sec.
  The running time of streaming on
  mapsMegamorphic is 3.7 sec, on filtersMegamorphic is 2.4 sec.
The evaluation platform is 1.8 GHz dualcore
Intel Core i5, 8 GB DDR3 main memory, macOS Big Sur 11.6.}
\label{f:bench}
\end{figure*}

We evaluated strymonas on the set of micro-benchmarks borrowed from
\cite{strymonas-2017}, to which we added ZipFilterFilter,
ZipFlatMapFlatMap (that is, zipping of two streams each containing a
filter, resp., flatmap operation) and runLengthDecoding
benchmarks. Fig.~\ref{f:bench} presents the results. Baseline is the
hand-written, hand-fused imperative (state-machine) code for the
entire pipeline, with no closures or thunks: what a competent
programmer would write to achieve best performance.
Streaming\footnote{\url{https://github.com/odis-labs/streaming}} is
one of the fastest streaming libraries in OCaml; it does not support
all the benchmark cases however. We also compare against Seq in the
OCaml standard library, and the iter library.

We have also implemented larger applications/benchmarks, such as FM
Radio reception. The generated C code, which compiles to fully
vectorized machine code, has the performance to sustain real-time
reception.

\bibliographystyle{plainnat}
\bibliography{../streams.bib}
\end{document}